\documentclass{article}
\usepackage{spconf,amsmath,graphicx}
\usepackage{epstopdf}
\usepackage{graphicx}
\usepackage{caption}
\usepackage{subfigure}
\usepackage{color}
\usepackage{placeins}
\usepackage{float}
\usepackage{tabularx,colortbl}
\usepackage{times,amsmath,epsfig}
\usepackage{epstopdf}
\usepackage{mathrsfs}
\usepackage{diagbox}
\usepackage{multirow}
\usepackage{algorithm}
\usepackage{algorithmic}
\usepackage{galois}
\usepackage{url}
\usepackage{amssymb}

\title{Perceptual Quality Study on Deep Learning based Image Compression}
\name{
Zhengxue Cheng$^{\star}$, Pinar Akyazi$^{\dagger}$, Heming Sun$^{\star}$, Jiro Katto$^{\star}$ and Touradj Ebrahimi$^{\dagger}$ }

\address{$^{\star}$ Department of Computer Science and Communication Engineering, Waseda University, Tokyo, Japan. \\
$^{\dagger}$ Multimedia Signal Processing Group (MMSPG), Ecole Polytechnique Federale \\
de Lausanne (EPFL), Lausanne, Switzerland.}

\begin{document}

\maketitle

\begin{abstract}
Recently deep learning based image compression has made rapid advances with promising results based on objective quality metrics. However, a rigorous subjective quality evaluation on such compression schemes have rarely been reported. This paper aims at perceptual quality studies on learned compression. First, we build a general learned compression approach, and optimize the model. In total six compression algorithms are considered for this study. Then, we perform subjective quality tests in a controlled environment using high-resolution images. Results demonstrate learned compression optimized by MS-SSIM yields competitive results that approach the efficiency of state-of-the-art compression. The results obtained can provide a useful benchmark for future developments in learned image compression.
\end{abstract}


\section{Introduction}
\label{sec:intro}

Image compression has been a popular research topic in the field of image processing for several decades. Conventional compression standards such as JPEG~\cite{IEEEexample:JPEG}, JPEG 2000~\cite{IEEEexample:JPEG2000}, and HEVC/H.265~\cite{IEEEexample:HEVC} rely on hand-crafted encoder-decoder (codec) architectures. They use fixed transforms such as discrete cosine transform (DCT) or discrete wavelet transform, in combination with with uniform quantization and entropy coder. Despite their substantial compression performance, the proliferation of high-resolution images and the development of novel image formats demand improvements on the existing compression algorithms towards an optimal solution for all types of image content.

Deep learning has been successfully applied to image compression in recent years~\cite{IEEEexample:waveone}-\cite{IEEEexample:CLIC} in which the autoencoder architecture is among those considered. An encoder-decoder pipeline based on autoencoder is claimed to provide efficient compression efficiency. Several approaches use generative adversarial training to achieve extremely low rates in~\cite{IEEEexample:waveone, IEEEexample:MITgan, IEEEexample:Extreme}. Variants are proposed to offer scalable compression in~\cite{IEEEexample:Toderici01, IEEEexample:Toderici, IEEEexample:Nick}. More general approaches adopt convolutional autoencoder (CAE) network with differentiable quantization and entropy model to achieve end-to-end learning, as reported in~\cite{IEEEexample:Theis}-\cite{IEEEexample:CLIC}. Promising results have already been achieved~\cite{IEEEexample:Balle2}.

\begin{figure}[tb]
	\centerline{\psfig{figure=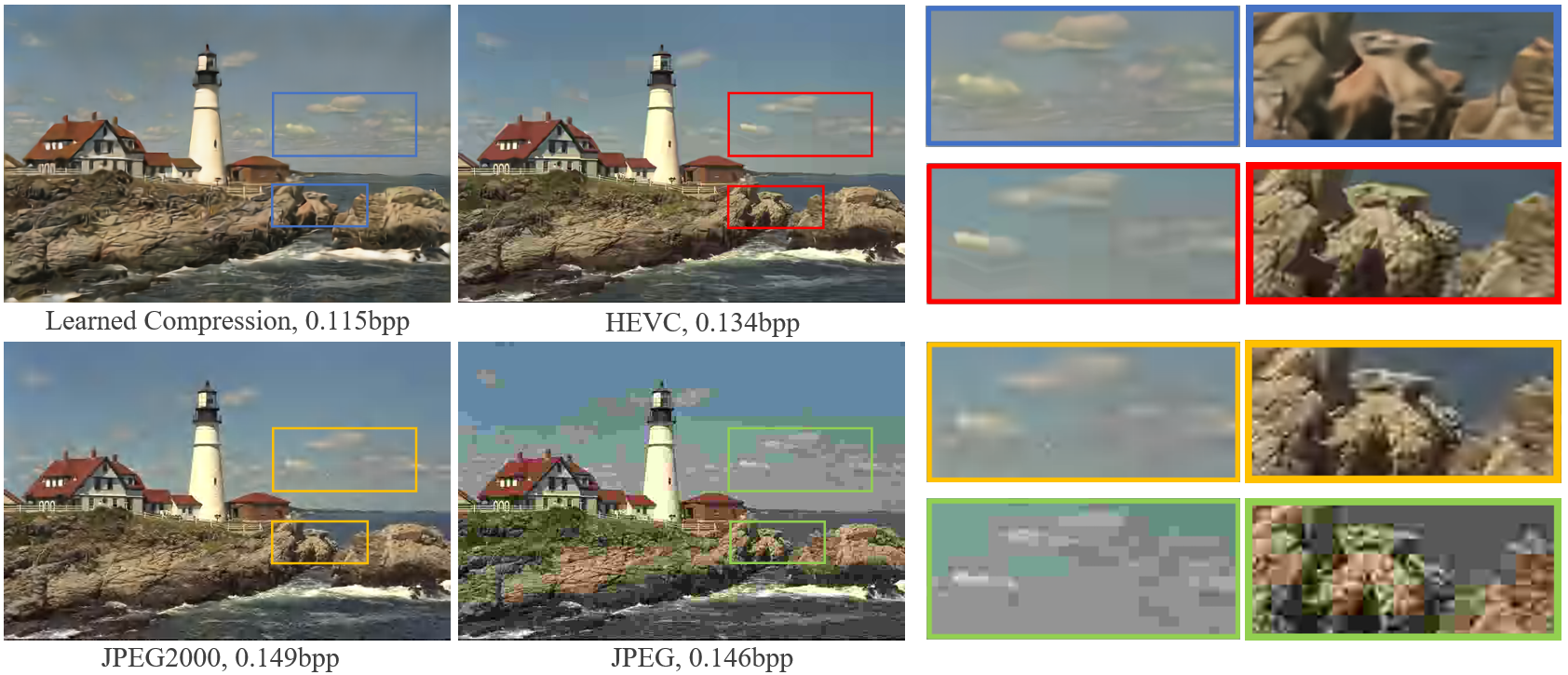,width=85mm} }
	\caption{Different artifacts caused by different compression algorithms. \emph{kodim21} from Kodak dataset is depicted above as an example.}
	\label{fig:visualization2}
\end{figure}

Most techniques only report their performance in terms of objective quality metrics, e.g. PSNR or MS-SSIM~\cite{IEEEexample:msssim}. Very few, however, have put emphasis on evaluating their performance based on subjective quality assessments. An example is the work in ~\cite{IEEEexample:MMSP} which relies on small images ($736\times960$) not necessarily fit for the current trend in high-resolution imaging applications. A particularly important observation is that learned compression brings new types of artifacts that differ from blocking or ringing artifacts created by traditional codecs, as illustrated in Fig.~\ref{fig:visualization2}. One observes that shape of clouds in the above illustration tend to be well-preserved in learned compression while clear artifacts can be seen in traditional coding approaches. However, the rock distorted by learned compression looks unnatural while the rock reconstructed by traditional codecs look more realistic. The impact of artifacts produced by learned image compression on human perception are still unknown. Therefore, a study on perceptual quality of learned image compression is essential in order to achieve further progress in this direction.

Our contributions in this paper are two-fold. First, we carefully design a generic learned image compression approach. Different objective quality losses are used to optimize our models as in many prior efforts in the sate of the art. Second, we conduct subjective quality assessments to compare the performance of six representative compression algorithms. Subjective quality evaluation results demonstrate that the learned compression algorithm optimized by MS-SSIM yields competitive results with state-of-the-art image compression. More importantly, we gain valuable insights on the future developments for learned compression.

\section{Codec Architecture}
\label{subsect.dataset}

\begin{figure}[tb]
	\centerline{\psfig{figure=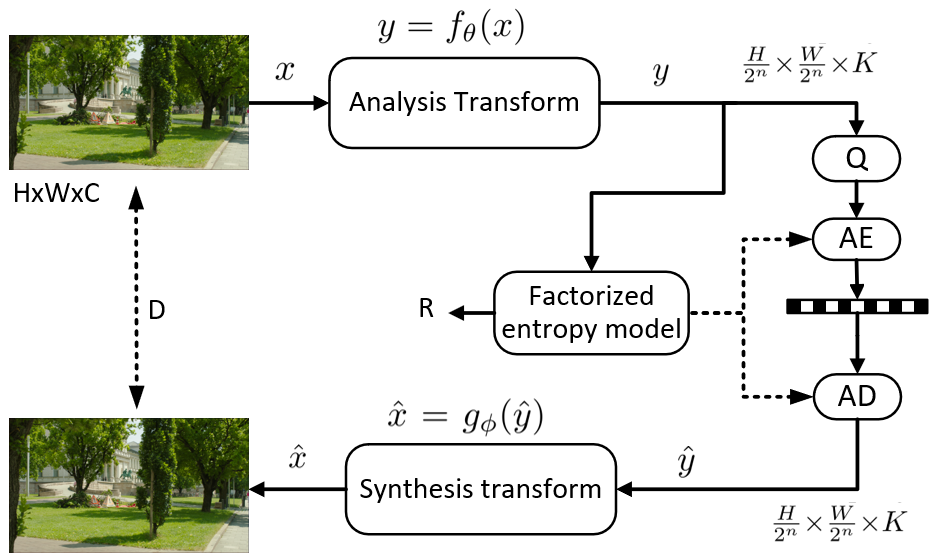,width=72mm} }
	\caption{Proposed learned image compression approach.}
	\label{fig:overall}
\end{figure}

We build a general learned compression based on a convolutional autoencoder (CAE)~\cite{IEEEexample:PCS}, shown in Fig.~\ref{fig:overall}, where Q represents quantization, and AE and AD represent the arithmetic encoder and arithmetic decoder, respectively. The analysis and synthesis transforms can be decomposed into downsampling and upsampling units, where a downsampling unit is composed of two convolution filters. The upsampling unit has the same structure with the convolution filters replaced by deconvolution filters. Each filter has a kernel size of 3 and 128 output channels. The latent representation $y$ has the dimension of $\frac{H}{2^{n}}\times\frac{W}{2^{n}}\times{K}$ where the $n$ is the number of down(up)sampling units and $K$ is the number of channels before quantization. In our experiments, we set $n=3$, $K=48$. We use factorized entropy model~\cite{IEEEexample:Balle2}, which is proved to be efficient to model any arbitrary distribution. It produces a context model and generates an estimated entropy to serve for AE and AD. For testing, we use the JPEG 2000 entropy coder to generate compressed bitstreams.

The loss function is defined similar to rate-distortion optimization (RDO) in traditional codecs, defined by
\begin{equation}
\label{eq.1}
J(\theta, \phi; x) = \lambda D(x, \hat{x})+ R(\hat{y})
\end{equation}
where $\lambda$ controls the trade off between the rate and distortion. $R$ represents the number of bits to encode the quantized compressed data $\hat{y}$. $\theta$ and $\phi$ are optimized parameters at the encoder and decoder sides. $D$ represents the distortion between original $x$ and reconstructed image $\hat{x}$, and can be estimated by any objective quality metrics. The reconstruction quality of learned compression heavily relies on the quality metrics in the loss function. The two most popular ones are
\begin{equation}
D(x, \hat{x}) = (1 - \text{MS-SSIM}(x, \hat{x}))
\end{equation}
or
\begin{equation}
D(x, \hat{x}) = \frac{1}{n}\sum^{n}_{i=1}(x-\hat{x})^{2}
\end{equation}
The model was optimized using Adam~\cite{IEEEexample:adam} with a batch size of 16 up to $10^{6}$ iterations. The learning rate was set at a fixed value of $1\times10^{-4}$. We have tested models trained using MSE and MS-SSIM loss to investigate the effect of quality metrics.

\begin{figure}[tb]
	\centerline{\psfig{figure=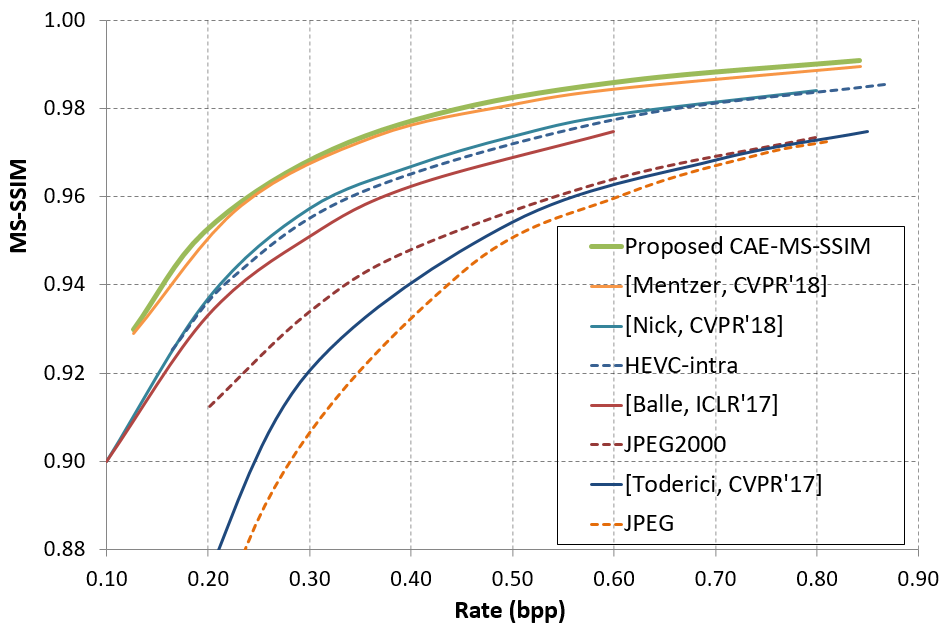,width=62mm} }
	\caption{Performance of recent works on Kodak dataset.}
	\label{fig:perf}
\end{figure}

In order to cope with high-resolution images, they are split into tiles and each coded  individually. After decoding, tiles are either stitched together as they do not overlap, or combined by weighted averaging of their boundary regions that overlap. We used a 32 pixel overlap as a compromise between redundancy and reduction of blocking artifacts between tiles. Fig.~\ref{fig:perf} depicts the performance of the proposed autoencoder codec in terms of MS-SSIM when compared to other state-of-the-art codecs on Kodak dataset. 

The most recent JPEG XL call for proposals~\cite{IEEEexample:JPEGXL} was used for anchor generation and resulted in a total of 6 codecs to be considered as in Table~\ref{Table.schemes}.

\begin{table*}[bt]
\begin{center}
\caption{Codecs considered in this paper.}
\label{Table.schemes}
\begin{tabular}{|l|l|}
 \hline
 \textbf{Codec} & \textbf{Specification and reference software}\\
 \hline
 CAE-MSE-ov          & Convolutional autoencoder optimized by MSE, stitched in an overlapped manner \\
 CAE-MS-SSIM-nonov    & Convolutional autoencoder optimized by MS-SSIM, stitched in a non-overlapped manner\\
 CAE-MS-SSIM-ov       & Convolutional autoencoder optimized by MS-SSIM, stitched in an overlapped manner \\
 HEVC/H.265           & ISO/IEC 23008-2|ITU-T Rec. H.265, Software: HM16.18+SCM-8.7, Intra~\cite{HM}\\
 JPEGXT               & ISO/IEC 18477, Software: JPEG XT v1.53~\cite{JPEGXT} \\
 JPEG2000             & ISO/IEC 15444-1|ITU-T Rec. T.800, Software: Kakadu v7.10.2~\cite{kakadu}\\
 \hline
\end{tabular}
\end{center}
\end{table*}

\section{Subjective Quality Evaluations}

\subsection{Dataset}

\begin{figure*}[bt]
\centering
\subfigure[APPLE]{
\label{Fig.sub.1}
\includegraphics[height=18mm]{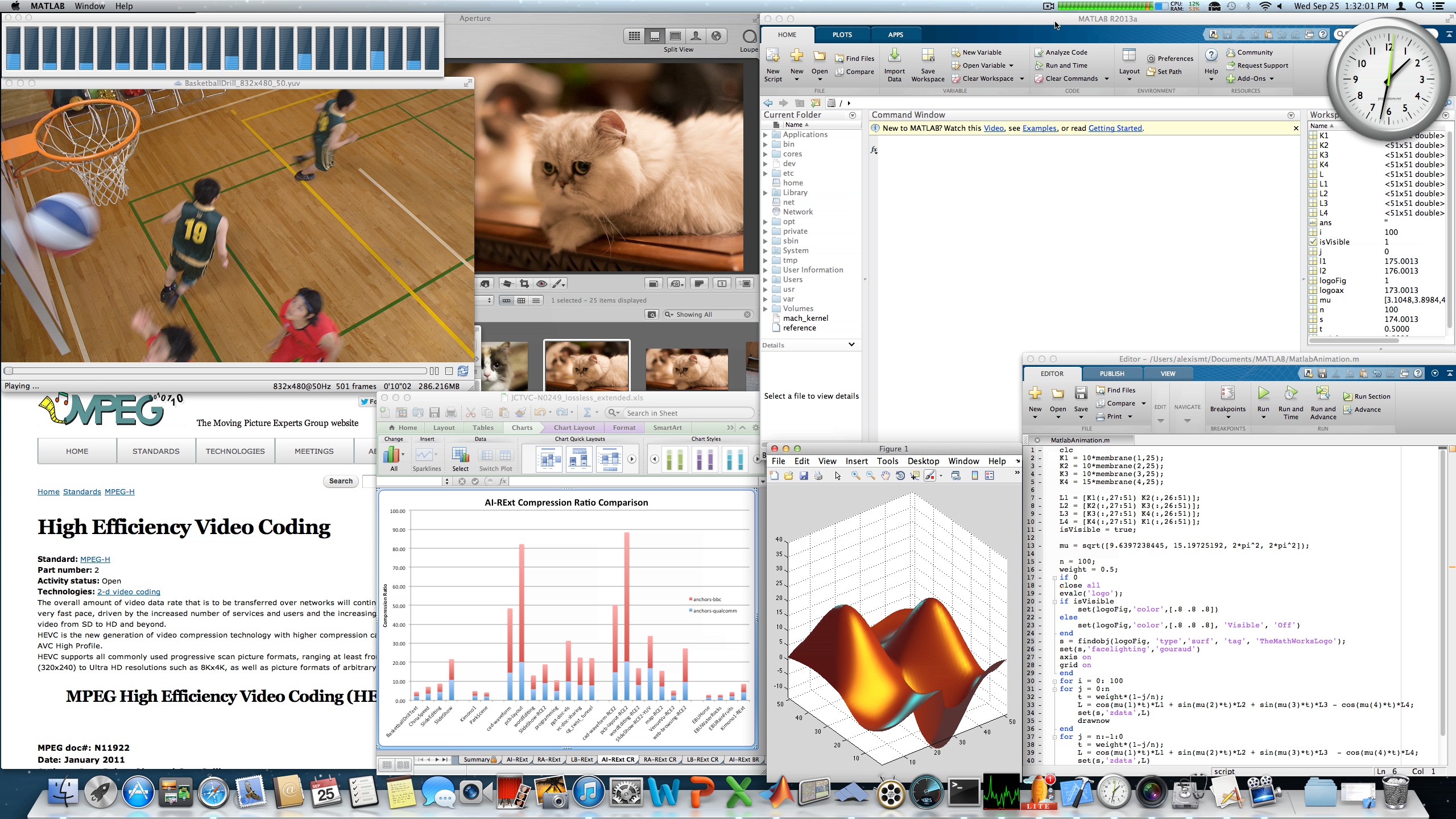}}
\subfigure[ARRI]{
\label{Fig.sub.3}
\includegraphics[height=18mm]{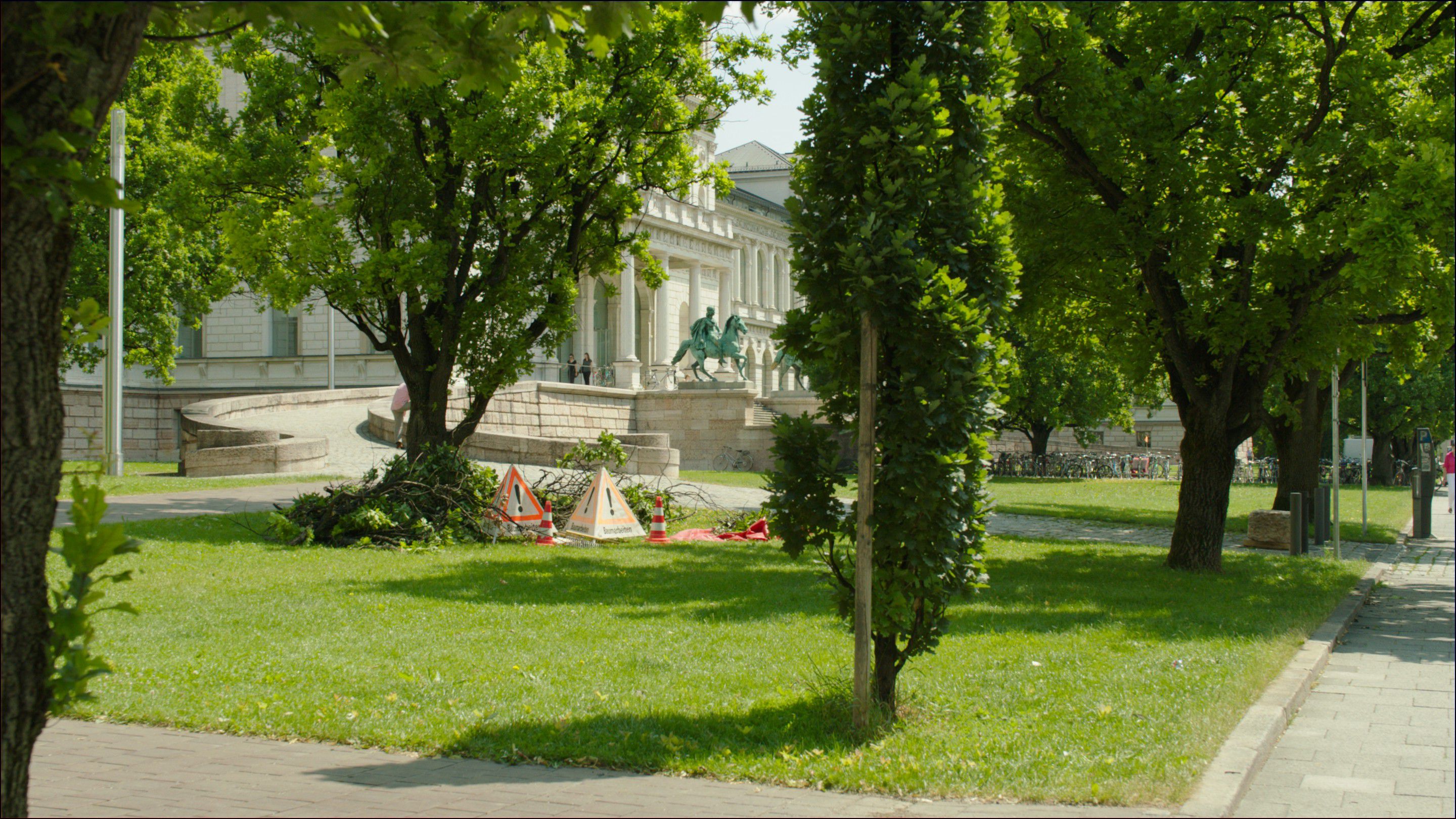}}
\subfigure[BIKE]{
\label{Fig.sub.4}
\includegraphics[height=18mm]{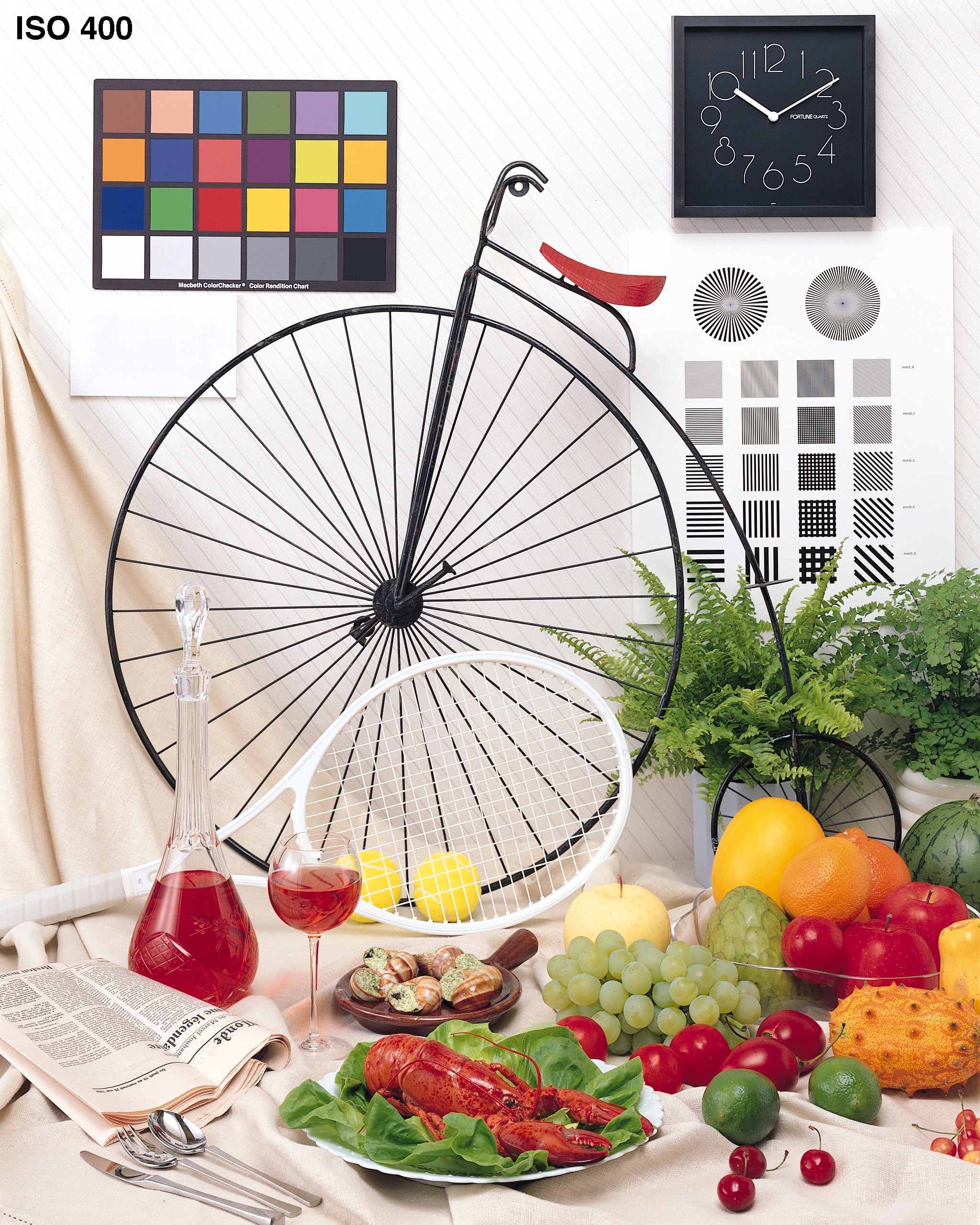}}
\subfigure[CAFE]{
\label{Fig.sub.2}
\includegraphics[height=18mm]{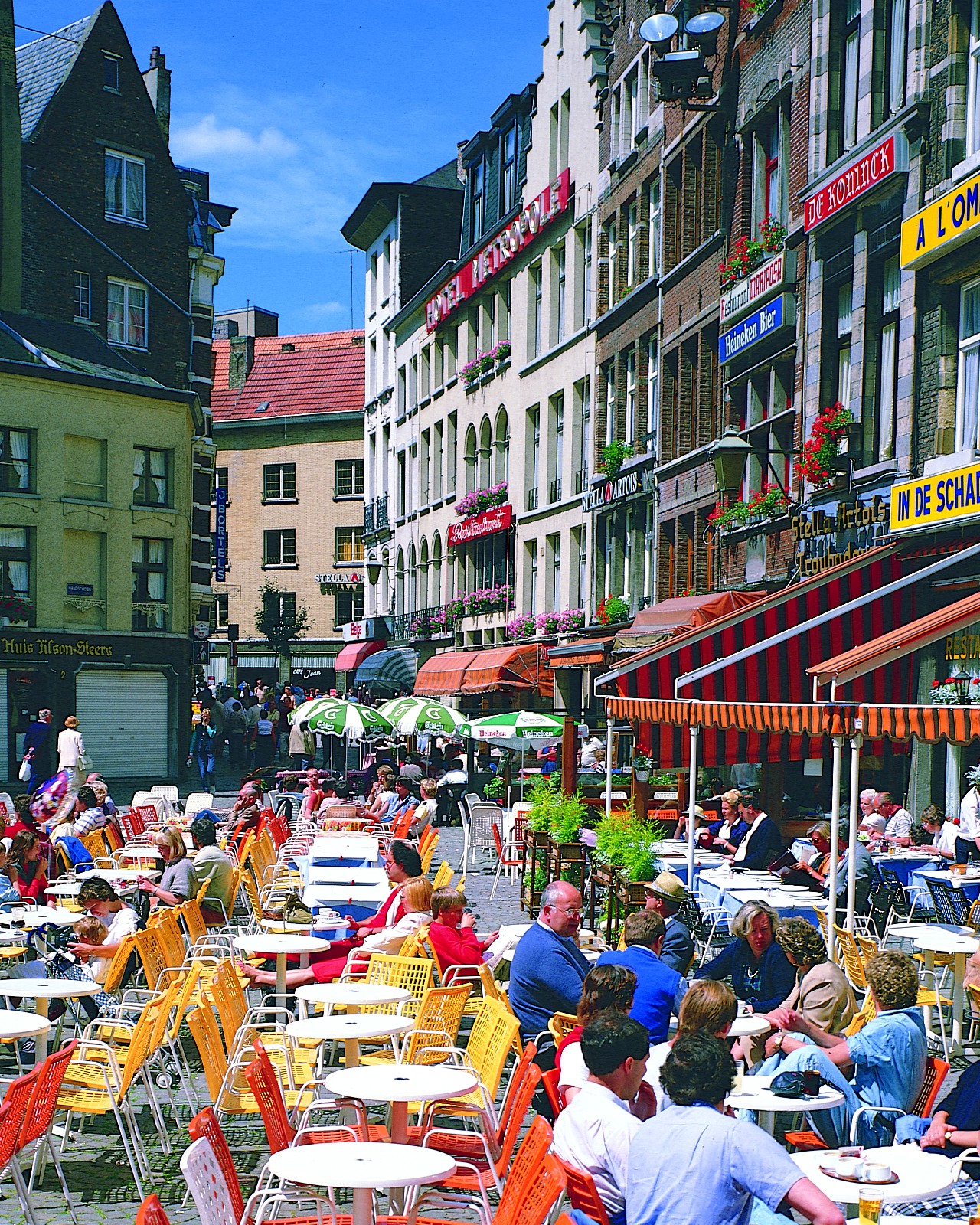}}
\subfigure[FemaleStripedHorseFly]{
\label{Fig.sub.1}
\includegraphics[height=18mm]{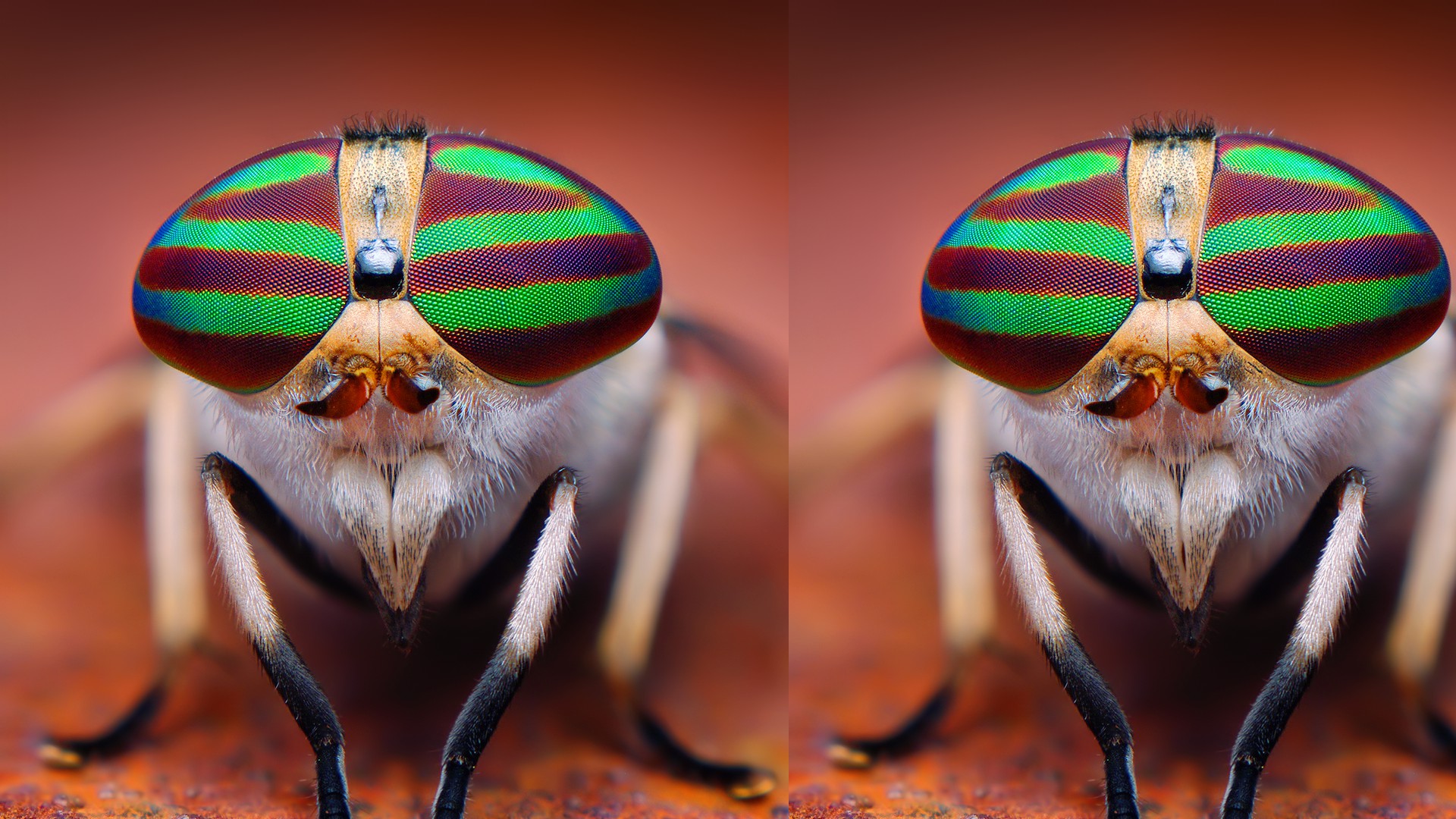}}
\subfigure[P06]{
\label{Fig.sub.1}
\includegraphics[height=18mm]{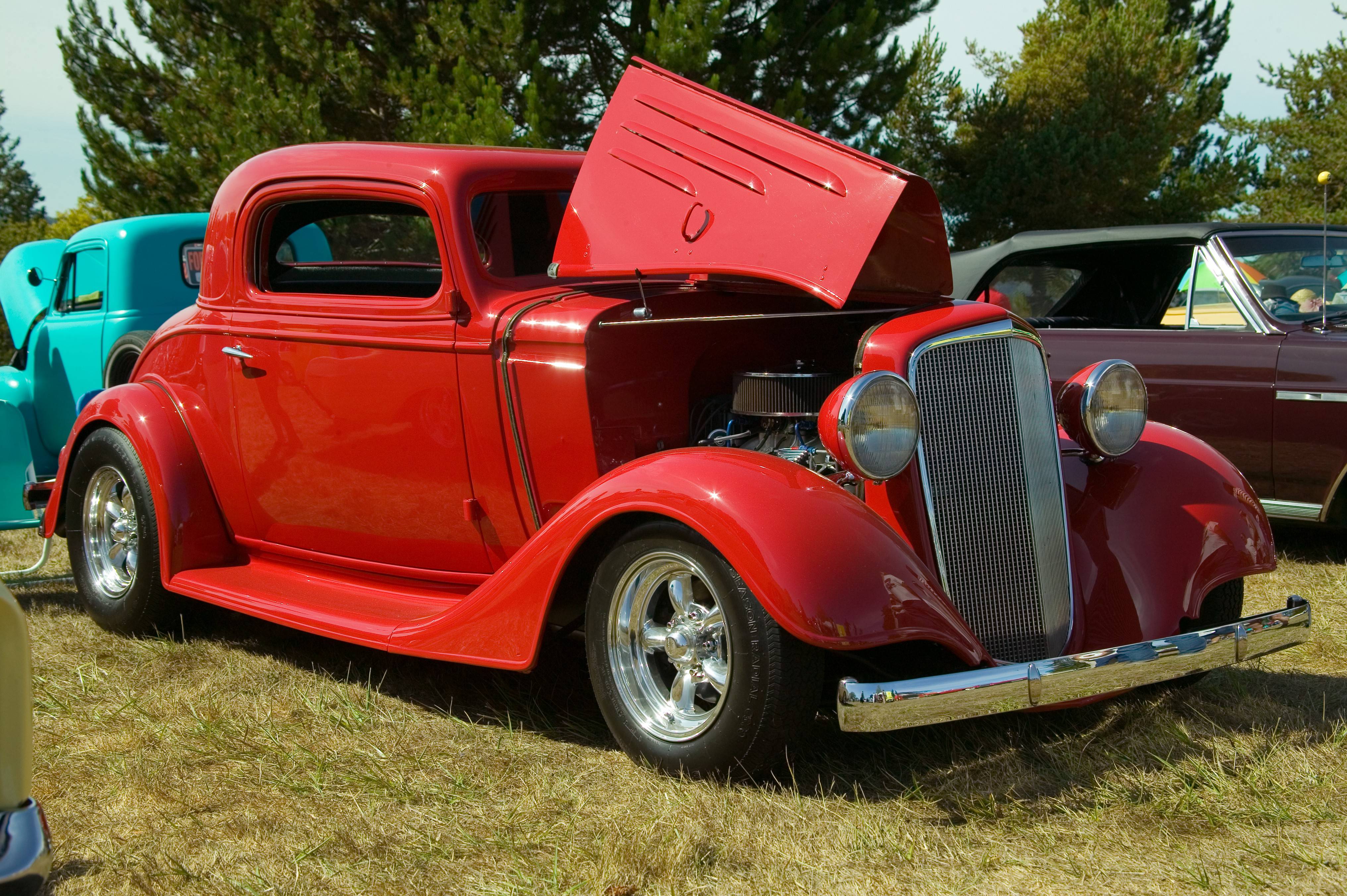}}
\subfigure[Woman]{
\label{Fig.sub.1}
\includegraphics[height=18mm]{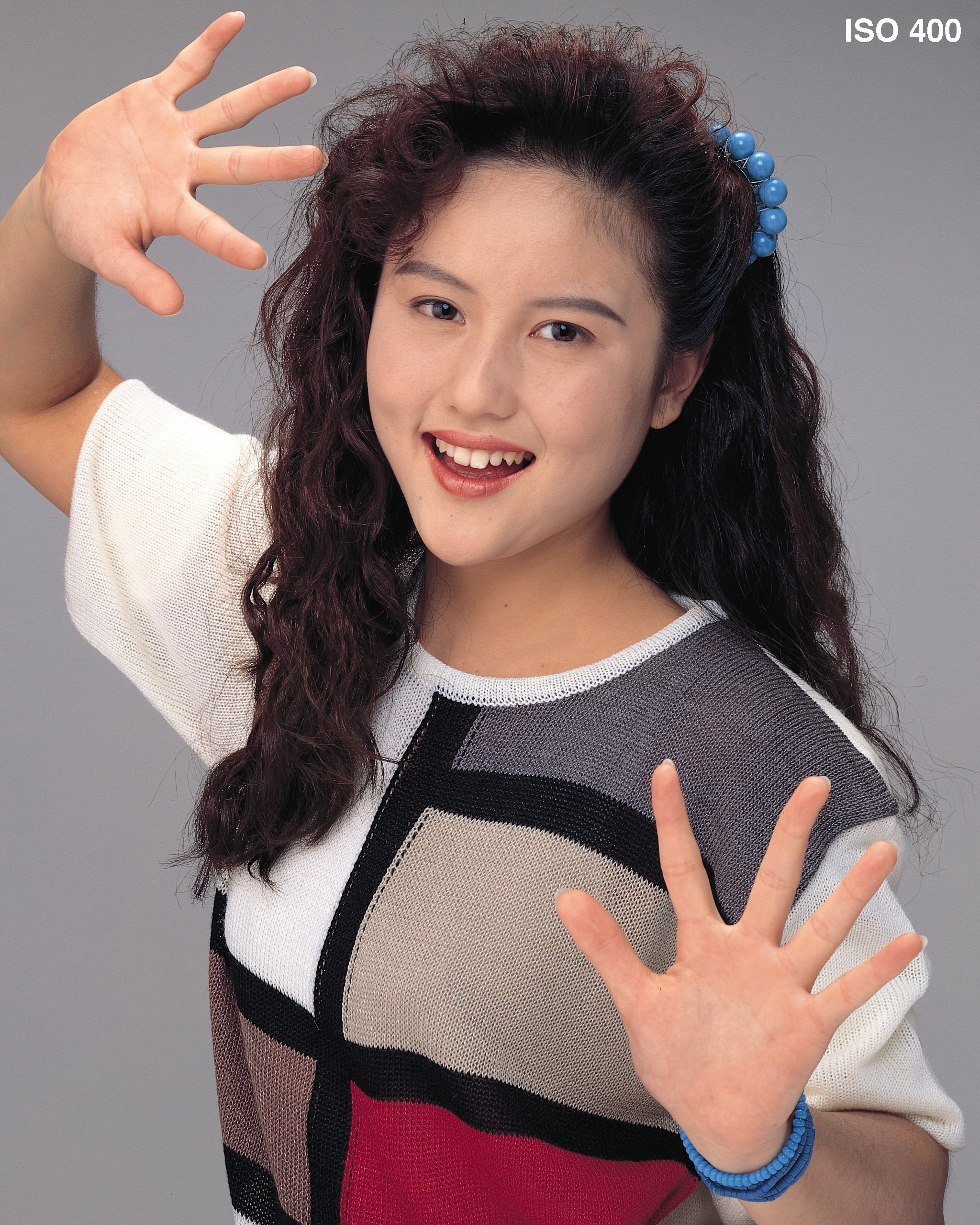}}
\caption{Test images in this study. }
\label{fig:testimages}
\end{figure*}

For this study, we selected 7 uncompressed 8-bit RGB test images in high-resolutions (HD to UHD), following the latest JPEG XL Call for Proposals~\cite{IEEEexample:JPEGXL}, as shown in Fig.~\ref{fig:testimages}. The codecs were evaluated on four target bitrates $R_1$ to $R_4$, corresponding to very low to high bitrates, which were determined during expert sessions as described in~\cite{IEEEexample:JPEGXL}. For image compression standards, we selected quantization parameters (QP) to match the target bitrates. For our proposed learned compression scheme, we adjust $\lambda$ in Eq.(\ref{eq.1}) to train models and achieve target bitrates.

\subsection{Test Methodology}

The methodology is based on Absolute Category Rating with Hidden Reference (ACR-HR)~\cite{P.910} where only one image is displayed in the center of the screen at a time. Participants were required to rate the visual quality based on a five-level scale, i.e., Excellent (5), Good (4), Fair (3), Poor (2), Bad (1). The whole evaluation consisted of 168 stimuli, namely 6 codecs, 7 contents and 4 bitrates. The display order was randomized so that the same content was never displayed consecutively. The test was split into two sessions to avoid subjects fatigue. Prior to testing, a training was performed to familiarize subjects with the typical artifacts and the rating scale. 


To avoid the involuntary influence of external factors and to ensure the reproducibility of results, a controlled environment with mid-gray background for subjective quality assessment was preferred according to~\cite{BT.500}. An Eizo ColorEdge CG318-4K monitor with native resolution of 4096$\times$2160 pixels was used for tests. The background of the display was set to mid grey~\cite{BT.2022}. The display brightness was calibrated at 120 cd/m2 and background illumination was set to 15 lux. A total of 16 participants (10 males and 6 females) took part in experiments, with an age between 19 and 38 years old, with an average of 26.4 and a median of 27. 

\section{Results and Discussion}

\begin{figure*}[tb]
\centering
\subfigure[APPLE]{
\label{Fig.sub.1}
\includegraphics[height=32mm]{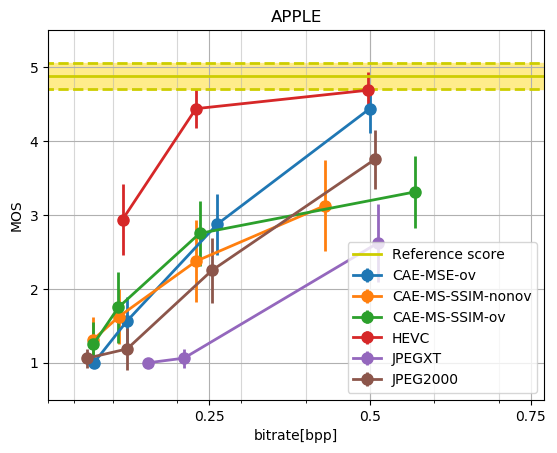}}
\subfigure[CAFE]{
\label{Fig.sub.2}
\includegraphics[height=32mm]{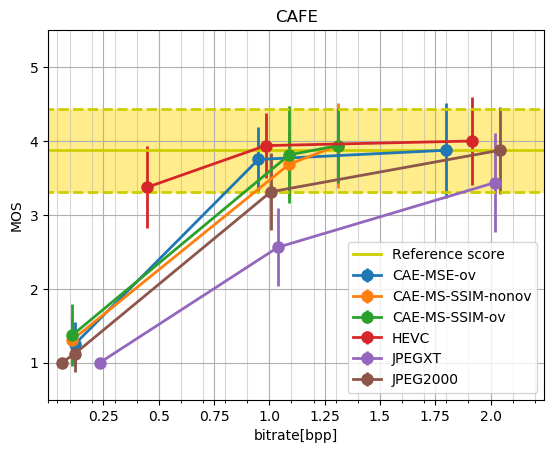}}
\subfigure[P06]{
\label{Fig.sub.1}
\includegraphics[height=32mm]{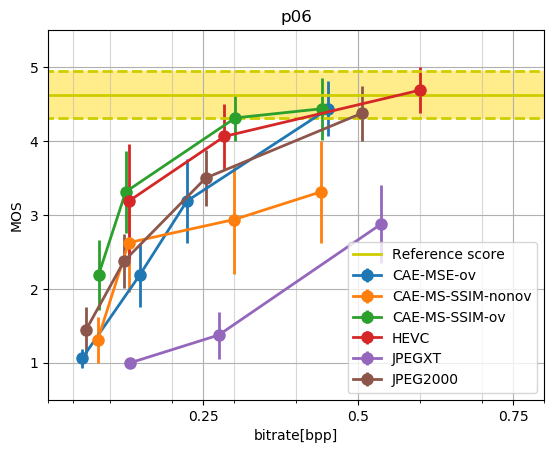}}
\subfigure[Woman]{
\label{Fig.sub.1}
\includegraphics[height=32mm]{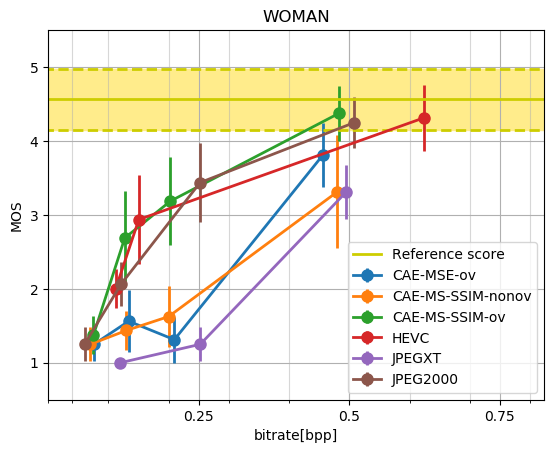}}
\caption{Results of MOS vs. bitrate with corresponding confidence interval. }
\label{fig:mos}
\end{figure*}

\begin{figure*}[hbt]
\centering
\subfigure[$R_{2}$]{
\label{Fig.sub.3}
\includegraphics[height=43mm]{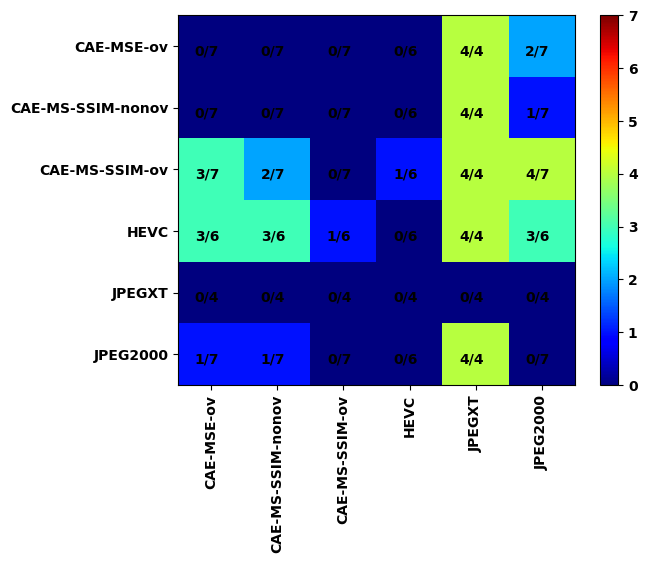}}
\subfigure[$R_{3}$]{
\label{Fig.sub.4}
\includegraphics[height=43mm]{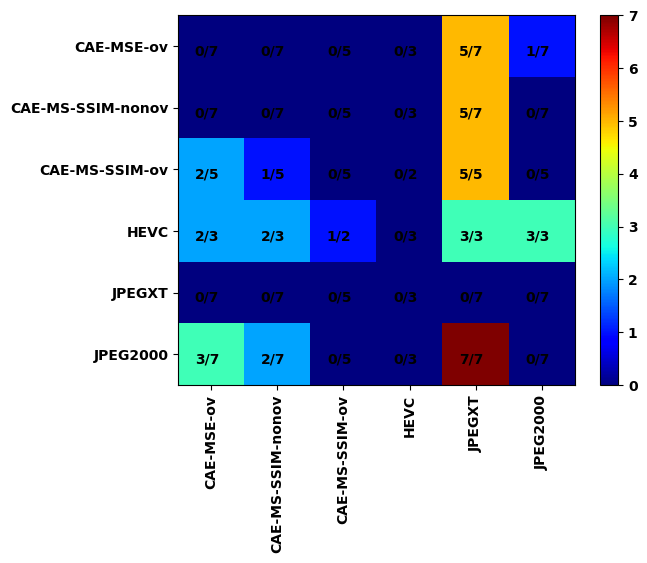}}
\subfigure[$R_{4}$]{
\label{Fig.sub.2}
\includegraphics[height=43mm]{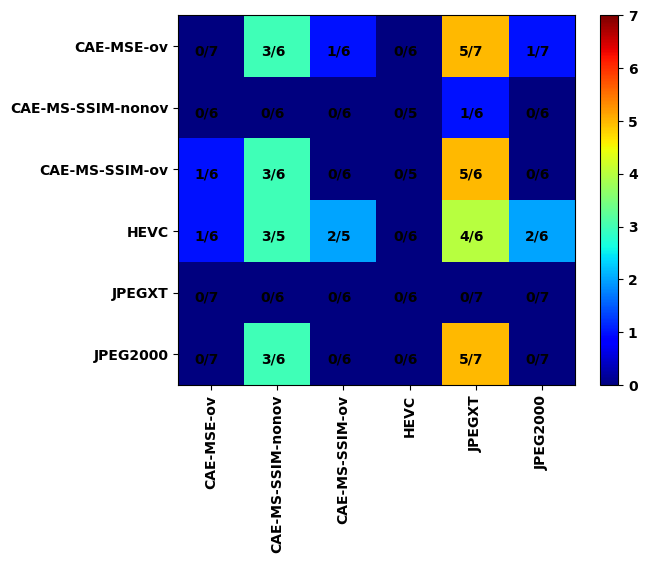}}
\caption{Pairwise Comparison for each bitrate, where for ``n/m'' in each cell, m denotes how many contents are comparable for each pairwise comparison, and n denotes how many times the codec on y-axis outperforms the codec on x-axis.}
\label{fig:paircomp2}
\end{figure*}

For the evaluation of perceived quality, outlier detection was performed using the approach in~\cite{BT.500} and no outlier was detected. The mean opinion score (MOS) was computed for each stimulus as
\begin{equation}
MOS_{j} = \frac{1}{N}\sum_{i=1}^{N}m_{ij}
\end{equation}
where $m_{ij}$ is the score for stimulus j given by subject $i$ and $N$ is the total number of participants. 95\% confidence intervals (CIs) were computed assuming a Student's t-distribution of the scores. Fig.~\ref{fig:mos} shows the MOS vs. bitrate curves for 4 typical contents. MOS of the hidden reference with corresponding CI are depicted with a yellow stripe, whereas all the codecs are plotted with a solid line. To determine whether the results yield statistical significance, a two-sided Welch test at 5\% significance level was performed on the scores.


Fig.~\ref{fig:paircomp2} shows for how many contents the codec on the y-axis performs significantly better than the codec on the x-axis for each bitrate. The minimum value is 0 and the maximum value is 7, corresponding to the total number of test contents. Some of the codecs could not reach all target bitrates (especially the lowest) within reasonable deviation due to limited and integer quantization parameter. To ensure fair comparison, we excluded the lowest bitrates and conducted the pairwise comparison only when the bitrate difference between actual bitrate and target bitrate was less than a predefined threshold specific to each targeted bitrate.

Fig.~\ref{fig:paircomp2} shows that HEVC/H.265-intra outperforms all codecs for all bitrates and all contents except for content \emph{Woman} at $R_2$, where CAE-MS-SSIM-ov outperforms HEVC/H.265-intra. This implies a better reconstruction quality for learned image compression approach for face contents. This result is expected as optimizing MS-SSIM leads to better structural similarity and human visual system is sensitive to human faces. Overall, CAE-MS-SSIM-ov achieves the second best performance, as it is statistically comparable with HEVC/H.265-intra at $R_2$, but a little worse at $R_3$ and $R_4$.

We can also conclude some useful information for learned image compression. First, comparing CAE-MS-SSIM-ov and CAE-MS-SSIM-nonov shows that the performance of the overlapping stitching strategy is superior to the non-overlapping case. More interestingly, CAE-MS-SSIM-ov outperforms CAE-MSE-ov on 3, 2 and 1 out of 7 contents at $R_2$, $R_3$ and $R_4$, respectively. On the contrary, CAE-MSE-ov outperforms CAE-MS-SSIM-ov only on 1 out of 6 contents at $R_4$. We have observed that to achieve higher subjective quality at low bitrates, MS-SSIM works better than MSE, while at high bitrates, MS-SSIM and MSE do not show significant differences.

\begin{figure}[tb]
\centering
\subfigure[MS-SSIM]{
\label{Fig.sub.1}
\includegraphics[height=31mm]{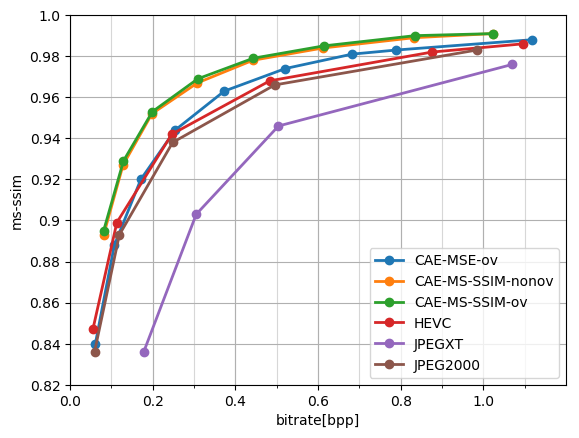}}
\subfigure[PSNR]{
\label{Fig.sub.3}
\includegraphics[height=31mm]{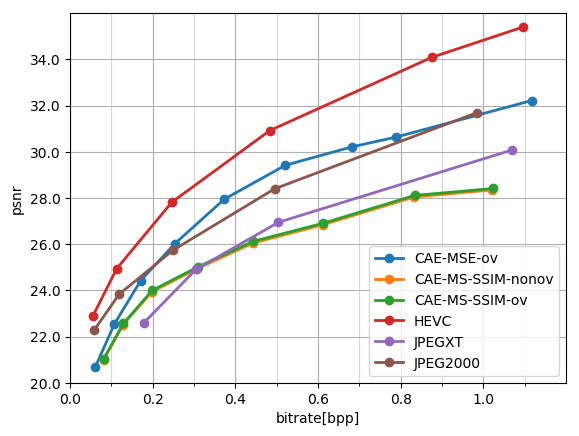}}
\caption{Performance with respect to MS-SSIM and PSNR. }
\label{fig:obj}
\end{figure}

Furthermore, we calculate the PSNR and MS-SSIM results to illustrate the difference between subjective and objective quality evaluations. To obtain a fair comparison, we use averaged PSNR on RGB components, defined by
\begin{equation}
\text{PSNR} = 10\log_{10}\bigg(\frac{255^{2}\times3}{MSE_{R} + MSE_{G} + MSE_{B}}\bigg)
\end{equation}
The MS-SSIM calculation refers to~\cite{IEEEexample:msssim}. Fig.~\ref{fig:obj} shows the averaged results on 7 contents. It can be observed that CAE-MS-SSIM-ov achieves the best MS-SSIM performance, which outperforms HEVC/H.265-intra significantly. In terms of PSNR, CAE-MSE is comparable to JPEG 2000, but worse than HEVC/H.265-intra. Both results correlate poorly with subjective qulaity evaluation results, which illustrates the need for a better perceptual similarity metric to improve the performance of learned based compression.

\section{Conclusion}

This paper presents a perceptual quality assessment study on the performance of learned image compression. First, we introduced a generic learned image compression approach and optimized it with two separate quality metrics. We considered (non)overlapping tiles to tackle memory limitations for high-resolution images. A total of six compression algorithms were involved in subjective quality assessment tests and high-resolution images were selected carefully, in line with state-of-the-art codec comparisons. We then conducted the subjective tests in a controlled environment by adopting an ACR-HR methodology. Results demonstrate learned compression optimized by MS-SSIM achieved competitive results with state-of-the-art codecs. We also observed the advantages of optimization with respect to MS-SSIM at low bitrates when compared to PSNR for learned image compression. Our study presents a thorough quality evaluation methodology that correlates well with human subjects opinion.


\end{document}